\documentclass[english,preprint,12pt]{elsarticle}
\usepackage[T1]{fontenc}
\usepackage[utf8]{inputenc}
\usepackage{amsmath}
\usepackage{amssymb}
\usepackage{verbatim}
\usepackage{graphicx}
\usepackage{esint}
\usepackage{babel}
\usepackage{color}
\usepackage[normalem]{ulem}
\journal{Physica A}

\begin{document}
\begin{frontmatter}

\title{Tsallis entropy approach to radiotherapy treatments}

\author[ad1,ad2]{O. Sotolongo-Grau\corref{tyf}}
\cortext[tyf]{Corresponding author, Tlf: +34 652575178, Fax: +34 934101701}
\ead{osotolongo@fundacioace.com}

\author[ad2]{D. Rodriguez-Perez}
\ead{daniel@dfmf.uned.es}

\author[ad3]{O. Sotolongo-Costa}
\ead{osotolongo@fisica.uh.cu}

\author[ad2,ad3]{J. C. Antoranz}
\ead{jcantoranz@dfmf.uned.es}

\address[ad1]{Fundació ACE, Institut Català de Neurociències Aplicades, 08029 Barcelona, Spain}
\address[ad2]{UNED, Departamento de Física Matemática y de Fluidos, 28040 Madrid,
Spain}
\address[ad3]{University of Havana, Cátedra de Sistemas Complejos Henri Poincaré,
Havana 10400, Cuba}

\begin{abstract}
The biological effect of one single radiation dose on a living tissue
has been described by several radiobiological models. However, the
fractionated radiotherapy requires to account for a new magnitude:
time. In this paper we explore the biological consequences posed by
the mathematical prolongation of a previous model to fractionated treatment.
Nonextensive composition rules are introduced to obtain the survival
fraction and equivalent physical dose in terms of a time dependent
factor describing the tissue trend towards recovering its radioresistance
(a kind of repair coefficient). Interesting (known and new) behaviors
are described regarding the effectiveness of the treatment which is
shown to be fundamentally bound to this factor. The continuous limit,
applicable to brachytherapy, is also analyzed in the framework of
nonextensive calculus. Also here a coefficient arises that rules the
time behavior. All the results are discussed in terms of the clinical
evidence and their major implications are highlighted.
\end{abstract}

\end{frontmatter}

\section{Introduction}

As can be seen in \cite{Swinney20041}
(and other works in the same issue) nonextensive Tsallis entropy 
\cite{tsallis-1999-29} has become a successful tool to describe a 
vast class of natural systems. A recently developed model 
\cite{Sotolongo_prl2010} of radiobiology shows 
this entropy definition could also be applied, not only to 
the development of living systems \cite{hydra1, hydra2}, but also to radiotherapy treatments.

The new radiobiological model (maxent model
in what follows) takes advantage of Tsallis entropy expression to describe
the survival fraction as a functional of the radiation absorbed dose. 
This model is also based on a minimum number of statistical 
and biologically motivated hypotheses.

The maxent model assumes the existence of a critical dose, $D_{0}$,
that annihilates every single cell in the tissue. The radiation dose
can be written as a dimensionless quantity in terms of that critical
dose as $x=d/D_{0}$, where $d$ is the radiation dose. Then the support
of the cell death probability density function, $p(x)$, in terms
of the absorbed dose $x$, becomes $\Omega=\left[0;1\right]$. Tsallis
entropy functional can be written,
\begin{equation}
S_{q}=\frac{1}{q-1}\left[1-\int_{0}^{1}p^{q}\left(x\right)dx\right],\label{eq:qentropy}
\end{equation}
where $q$ is the nonextensivity index. The survival fraction of cells
will be given by $f\left(x\right)=\int_{x}^{1}p\left(x\right)dx$,
that is the complement of the fraction of cells killed by radiation.
In order to maximize functional \eqref{eq:qentropy} we must consider
the normalization condition,
\begin{equation}
\int_{0}^{1}p\left(x\right)dx=1\label{eq:norm}
\end{equation}
 Also, following \cite{Plastino99}, we must assume the existence
of a finite $q$-mean value, 
\begin{equation}
\int_{0}^{1}p^{q}\left(x\right)xdx=\left\langle x\right\rangle _{q}\label{qmean}
\end{equation}
Then the Lagrange multipliers method leads to, 
\begin{equation}
p\left(x\right)=\gamma\left(1-x\right)^{\gamma-1},\label{eq:px}
\end{equation}
with $\gamma=\frac{q-2}{q-1}$. So, the survival fraction predicted by the model
is 
\begin{equation}
f\left(x\right)=(1-x)^{\gamma},\label{eq:fx}
\end{equation}
valid for $x\in\Omega$ and requiring $\gamma>1$. 

This model has shown a remarkable
agreement with experimental data \cite{Sotolongo_prl2010,Sotolongo10_chamonix}, 
even in those limits where previous models are less accurate,
mainly at high doses. The analysis of the model fit to experimental data also provides new
hints about the tissue response to radiation: first, the interaction
of a tissue with the radiation is universal and characterized by a
single exponent (not dependent on the radiation type, energy or dose rate); second,
the model includes a cutoff dose (this one, dependent on the characteristics of the radiation) above which every single
cell dies. Furthermore, previous models can be derived as particular
limiting cases. Finally, as for those models, its mathematical expression
is simple and can be easily plotted and interpreted.

The maxent model was derived for radiobiological 
survival fraction but its applicability could be extended
to other processes. Indeed, every phenomena 
describable in terms of Tsallis entropy \cite{Plastino1994140}, fulfilling the 
maximun entropy principle and exhibiting a critical cutoff (represented here by $x=1$),
must follow \eqref{eq:fx}. 

%In particular this could be applied to some 
%biological interactions or clinical treatments as antibiotics or 
%other killer drugs.

Nevertheless the expression \eqref{eq:fx}, understood as 
survival probability, lacks the extensivity property. In other words, 
for $n$ events following \eqref{eq:fx} the total survival probability 
should be found as a composition of the survival probabilities
of the successive events. However, there is not a straightforward 
composition rule for those probabilities.

Indeed, if two doses, $x_{A}$ and $x_{B}$ are applied, the resulting 
probability from their composition has two possible values. 
If the total dose is assumed additive, 
$f_{AB}=\left(1-x_{A}-x_{B}\right)^{\gamma}$, that is, the individual
probabilities  under $A$ and $B$ events 
could not be treated as independent probabilities, $f_{AB}\neq f_{A}f_{B}$.
On the other hand, if probabilities are multiplicative, 
$f=\left(1-x_{A}\right)^{\gamma}\left(1-x_{B}\right)^{\gamma}$
, doses would not fulfill the superposition principle for the equivalent
physical dose, $x_{AB}\neq x_{A}+x_{B}$. 

The subject of this manuscript is to develop on the composition rules
that would lead to the survival fraction and the equivalent physical
dose of a fractionated processes, and to derive the biological implications
of such rules. This will be approached within the frameworks of $q$-algebra and $q$-calculus
\cite{nivanen-2003-52,borges-2004-340,Kalogeropoulos2005408}, as
far as they are the natural ones for the maxent model.

\section{Composition rules}
Each event described by \eqref{eq:fx} represents a measured energy impact,
or dose, $x$ causing an irreversible effect or hazard
over a group of individual entities leading to a survival probability
of those entities. As \eqref{eq:fx}
represents a nonextensive process, an appropiate set of composition rules must 
be developed in order to find the effect of several combined events on the group
of entities.

As it has just been exposed, if those composition rules are defined
keeping the superposition principle for the dose, the probabilities
are not independent of each other and \emph{vice versa}, if the 
probabilities are multiplicative, the dose becomes 
non additive \cite{kaniadakis2005}.
Luckily, the nonextensive thermostatistics provides tools to 
find the right expressions in each case 
\cite{nivanen-2003-52,borges-2004-340,Kalogeropoulos2005408,kaniadakis2005}.

If the survival probabilities are independent, 
the total probability for two events $A$ and $B$ is $f_{AB} = f_{A}f_{B}$. 
So the nonextensive sum must be constructed as 
$x \oplus y = x + y - xy$ and we can write,
\begin{equation}
\begin{array}{c}
x_{AB} = x_{A} \oplus x_{B} = x_{A} + x_{B} - x_{A}x_{B}\\
f_{AB} = f_{A}f_{B}
\end{array}\label{eq:option1}
\end{equation}
On the other hand, if the dose is additive, $x_{AB} = x_{A} + x_{B}$, 
the nonextensive product must be 
$x \otimes y = \left( x^{1/\gamma} + y^{1/\gamma} -1 \right)^{\gamma}$ so we can write,
\begin{equation}
\begin{array}{c}
x_{AB} = x_{A} + x_{B}\\
f_{AB} = f_{A} \otimes f_{B} = \left( {f_{A}}^{1/\gamma} + {f_{B}}^{1/\gamma} -1 \right)^{\gamma}
\end{array}\label{eq:option2}
\end{equation}

The main issue here is that in clinical treatments both limits are not clearly distinct. 
Indeed, when events occur separate enough in time, tissue recovering capabilities
make physical consequences of one of them independent from the
others'. From a radiobiologist point of view this is similar to applying
the next radiotherapy session after late effects of the former occur.
However, if the events occur simultaneously the 
dose must be considered additive. In other words, \eqref{eq:option1} and \eqref{eq:option2}
represent limit cases of the interaction process corresponding 
to $t=\infty$ and $t=0$ respectively, where $t$ is the time between succesive events.

In order to describe a real fractionated process, 
new generalized sum and product operators need to be introduced, 
taking into account that \eqref{eq:option1} and \eqref{eq:option2} 
must hold in the multiplicative and additive limits, respectively.

The resulting probability in \eqref{eq:option1} is the product
of partial probabilities, and for the whole process,
\begin{equation}
F_{n}=\prod_{i=1}^{n}\left(1-x_{i}\right)^{\gamma},\label{eq:independiente}
\end{equation}
where $i$ runs along the events. 

However, if the dose is additive, the total survival
fraction follows,
\begin{equation}
F_{n}=\left(1-\sum_{i=1}^{n}x_{i}\right)^{\gamma}\label{eq:additive}
\end{equation}

Notice that it is possible to write
\eqref{eq:additive} as a product, finding the expression that turns
$F$ after $n-1$ events into $F$ after $n$ events. So, 
\eqref{eq:additive} can be recast in the form, 
\begin{equation}
F_{n}=\left(1-\frac{x_{n}}{1-\sum_{k=1}^{n-1}x_{k}}\right)^{\gamma}F_{n-1}=\prod_{i=1}^{n}\left(1-\frac{x_{i}}{1-\sum_{k=1}^{i-1}x_{k}}\right)^{\gamma}\label{eq:prod_redef}
\end{equation}
 This expression can be interpreted as a modified \eqref{eq:independiente}
in which the denominator, which plays the role of the annihilation
cutoff, gets reduced, in practice, by an amount $x_{i}$ after addition
of the $i$-th event. On the other hand, for independent events
this critical cutoff would remain constant along the whole process. 

The new operators for nonextensive sum, $\bigoplus$, and product,
$\bigotimes$, must be defined to hold,
\begin{equation}
F_{n}=\bigotimes_{i=1}^{n}\left(1-x_{i}\right)^{\gamma}=\left(1-\bigoplus_{i=1}^{n}x_{i}\right)^{\gamma}=\prod_{i=1}^{n}\left(1-\frac{x_{i}}{1-\epsilon\bigoplus_{k=1}^{i-1}x_{k}}\right)^{\gamma},\label{eq:final}
\end{equation}
subject to the condition $\bigoplus_{i=1}^{n}x_{i}\rightarrow\sum_{i=1}^{n}x_{i}$,
for $\epsilon\rightarrow1$. In this way, \eqref{eq:independiente} and 
\eqref{eq:prod_redef} will be the limits of the new operators. Indeed, 
the coefficient $\epsilon\in\left[0,1\right]$ acts as a session-coupling coefficient for 
equations \eqref{eq:independiente} and \eqref{eq:prod_redef}
such that $\epsilon=1$ implies events are completely
correlated while $\epsilon=0$ means they are fully independent, i.e. not coupled.

Even though \eqref{eq:final} gives
a closed and univocal definition of $\bigoplus$ and 
$\bigotimes$ operators, this is an implicit definition. In order to use these operators an 
explicit definition is desired.

The analytical expresion for the new 
operators $\oplus$ and $\otimes$ can be found assuming 
there is a single event with an effective dimensionless dose $X$
corresponding to the whole process such that,
\begin{equation}
F_{n}=\left(1-X\right)^{\gamma}=\left(1-\bigoplus_{i=1}^{n}x_{i}\right)^{\gamma}\label{eq:deffect_def}
\end{equation}
 After the $i$-th event, the dimensionless effective dose would become,
\begin{equation}
X_{i}=X_{i-1}+x_{i}\left(\frac{1-X_{i-1}}{1-\epsilon X_{i-1}}\right),\label{eq:recursive}
\end{equation}
 assuming $X_{1}=x_{1}$. When the $n$-th event is given, then $X_{n}=X$.

From this follows that,
\begin{equation}
\begin{array}{c}
x_{AB} = x_{A} \oplus x_{B}  = x_{A} + x_{B} \left( \frac{1-x_{A}}{1- \epsilon x_{A}} \right) \\
f_{AB} = f_{A} \otimes f_{B} = f_{A} \left[ \frac {{f_{B}}^{1/\gamma} - \epsilon \left(1- {f_{A}}^{1/\gamma} \right)}{1 - \epsilon \left(1- {f_{A}}^{1/\gamma} \right)}  \right]^{\gamma}
\end{array}\label{eq:operators}
\end{equation} and limit definitions \eqref{eq:option1} and \eqref{eq:option2} are recovered for $\epsilon=0$ and $\epsilon=1$ respectively. According
to both limit interpretations, session-coupling $\epsilon$ values will depend on the
time between events and also on tissue repair or recovery capabilities.

\section{Biological and physical implications}

\subsection{Isoeffect relationship}

%A single radiation fraction with an effective dimensionless dose $X$
%equal to the whole fractionated treatment can be found such that,
%\begin{equation}
%F_{n}=\left(1-X\right)^{\gamma}=\left(1-\bigoplus_{i=1}^{n}x_{i}\right)^{\gamma}\label{eq:deffect_def}
%\end{equation}
% After the $i$-th fraction, the dimensionless effective dose becomes,
%\begin{equation}
%X_{i}=X_{i-1}+x_{i}\left(\frac{1-X_{i-1}}{1-\epsilon X_{i-1}}\right),\label{eq:recursive}
%\end{equation}
% assuming $X_{1}=x_{1}$. When the $n$-th fraction is given, then $X_{n}=X$.
One of the central concepts of radiotherapy is isoeffect relationships. 
An oncologist usually seeks treatments that produce the best outcome on 
the target tumor ($F_{tumor}$), 
while causing at most the maximum allowed damage on the surrounding 
healthy tissues ($F_{tissue}$). 
In other words, he seeks among the pairs of values $(n, x)$ 
that give the same value of $F_{tissue}$ 
for the healthy tissue that one attaining
the maximum value of $F_{tumor}$. Given the expression 
\eqref{eq:deffect_def}  this can be reduced to find the pairs $(n, x)$ 
that render the same value of $X$.

Indeed, all fractionated treatments sharing the same value of effective dose, $X$, will provide
the same value for the survival fraction. So, the same $X$ will provide
the isoeffect criterion for the fractionated therapy.

\begin{figure}
\begin{centering}
\includegraphics[width=1\textwidth]{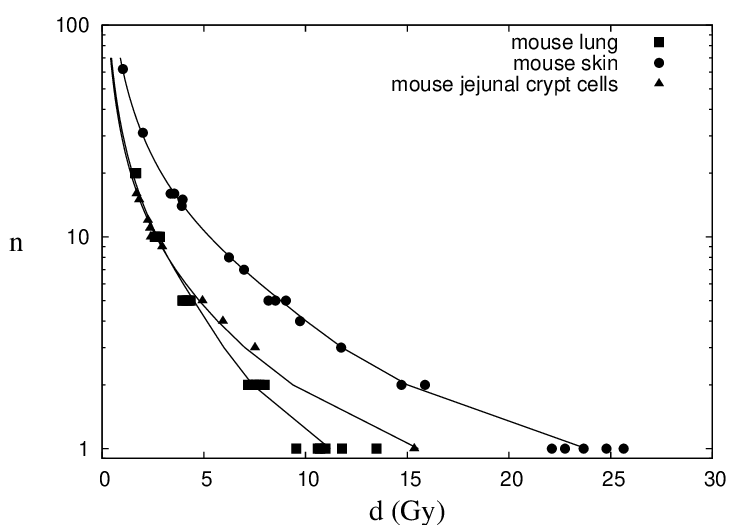} 
\par\end{centering}

\caption{\label{fig:Isoeffects-fits}Isoeffect relationship data reported for
mouse lung by \cite{isodata5} ($\epsilon=0.50$, $D_{0}=11.3\textrm{ Gy}$),
mouse skin by \cite{isodata3} ($\epsilon=0.58$, $D_{0}=24.0\textrm{ Gy}$)
and mouse jejunal crypt cells by \cite{isodata1} ($\epsilon=0.62$,
$D_{0}=16.1\textrm{ Gy}$), fitted to \eqref{eq:recursive}.}
\end{figure}

In order to check the model reliability, it has been fitted to data
from \cite{isodata5,isodata3,isodata1} using a weighted least squares
algorithm \cite{Sotolongo_prl2010}. Those data sets are considered as a reliable source of
clinical parameters (as the $\alpha/\beta$ relation of LQ model \cite{Steel_ch10}).
The results of the fit are shown in Figure \ref{fig:Isoeffects-fits}. 

%DRP: no se dice nada sobre el D_0 que da el ajuste; entiendo que puede
% despistar decir que  se encuentra en los rangos habituales encontrados
% en el PRL (supongo que un referee despistado preguntará no sólo eso,
% sino el valor del gamma)
%OSG: D_0 se menciona en la introduccion antes de adimensionalizar todo.
% Crei innecesario mencionarlo de nuevo.
The obtained session-coupling coefficients
show a survival fraction behavior far from the pure $q$-algebraic limits
($\epsilon=0$ and $\epsilon=1$). Since $\epsilon$ values for usual tissue
reaction differ from limiting values, it is worth to further study the
biophysical interpretation of this new parameter.

\begin{figure}
\begin{centering}
\includegraphics[width=1\textwidth]{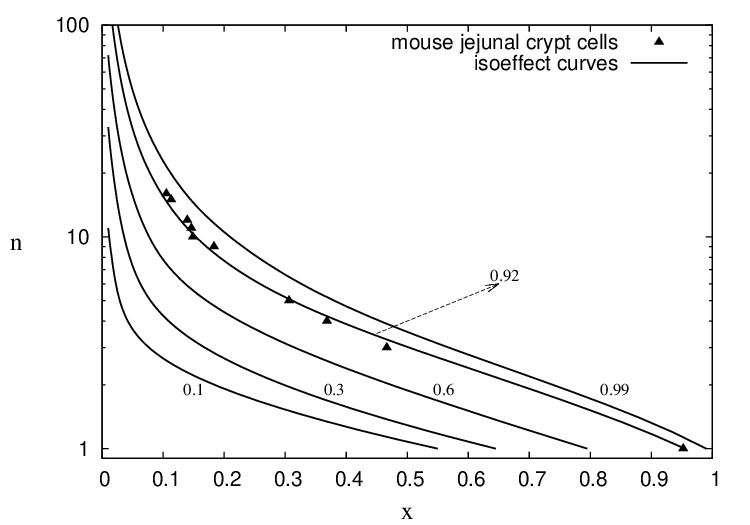} 
\par\end{centering}

\caption{\label{fig:Isoeffects-explain}Isoeffect curves for mouse jejunal
crypt cells by \cite{isodata1}. Curves are calculated based on fitted
parameters $\epsilon=0.62$ and $D_{0}=16.1\textrm{ Gy}$ for different
values of $X$ in \eqref{eq:recursive}, shown for every plot.}
\end{figure}

Every $X$ value provides a different isoeffect relationship, as shown
in figure \ref{fig:Isoeffects-explain}. Once
a treatment coefficient values ($\epsilon$ and $D_{0}$) are known, the dosage can be tuned
to obtain the desired effective dose by changing $n$ and $d$. 
Notice that $\gamma$ does not play any role in this composition, 
thus reducing the number of model parameters to take into account here.

%Esto de aqui es nuevo respondiendo a la cuarta pregunta del referee

As there is not enough experimental data available,
in order to find session-coupling $\epsilon$ values 
for known tissues or tumors we will use the LQ model of incomplete repairment 
to show how our model could be used to
assess the desired therapy schedule.

\begin{figure}
\begin{centering}
\includegraphics[width=1\textwidth]{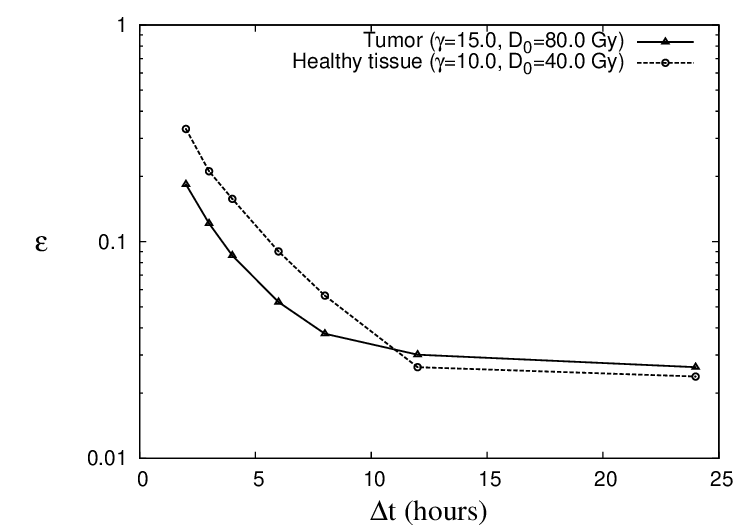} 
\par\end{centering}
\caption{\label{fig:epsilontime} Session-coupling values
found following the LQ model 
of incomplete repairment for an hypotetic tumor and healthy tissue
as function of treatment time schedule.}
\end{figure}

Let us suppose a healthy tissue
$H$ with $\gamma = 10.0$ and $D_{0}=40.0\textrm{ Gy}$ surrounding
a more resilient tumor $T$ with $\gamma = 15.0$ and
$D_{0}=80.0\textrm{ Gy}$. Now we will assume 
that $H$ can not receive more than $36.0 \textrm{ Gy}$ or $X=0.9$.
After finding the corresponding LQ model $\alpha$ and $\beta$ values is easy to reproduce 
the isoeffect curves for incomplete repairment following 
\cite{Steel_ch10} if the cell repair half time is known. 
We had chosen a repair half time of 3 hours for $H$ and $T$ but the same procedure 
could be applied for different repair half time values.
Each of these curves represents a different treatment
schedule characterized by the time ($\Delta t$) between sessions.
From these curves the $\epsilon$ values as a function of $\Delta t$ 
could be found as shown in Figure \ref{fig:epsilontime}.

\begin{figure}
\begin{centering}
\includegraphics[width=1\textwidth]{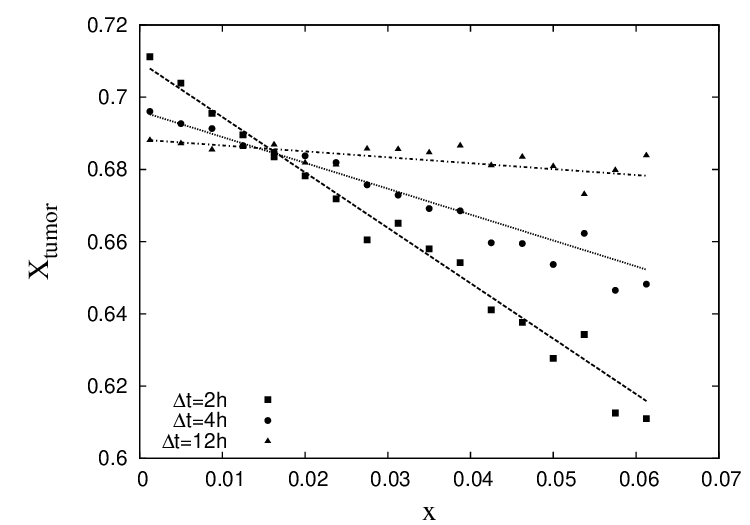} 
\par\end{centering}
\caption{\label{fig:xeftumor}$X$ values as function of session  
adimensional dose $x$
found for the hypotetic tumor $T$ following the treatment schedules 
of incomplete repairment as function of treatment time schedule. 
Lines represents the approximate
behaviour of $X$ values.}
\end{figure}

After the values of $\epsilon$ have been determined for the tumor 
then the effective dose $X$ received for each schedule could be found 
as shown in Figure \ref{fig:xeftumor}. This shows us that for small 
$x$ values the best outcome is reached at more consecutive
sessions, whereas for more separated sessions the appropriate dosage
is attained at higher $x$ values. In particular, for the case of sample 
tissues H and T, described above, best results are found with a more 
fractionated treatment with its fractions scheduled as close as possible.

Note that for a real example this procedure must be
followed after finding the experimental values
of $\epsilon$ for each schedule.
Even though illustrative, this example must be
taken with caution as it is based on another model whose
validity limits are not clear.

\subsection{Critical dosage}

\begin{figure}
\begin{centering}
\includegraphics[width=1\textwidth]{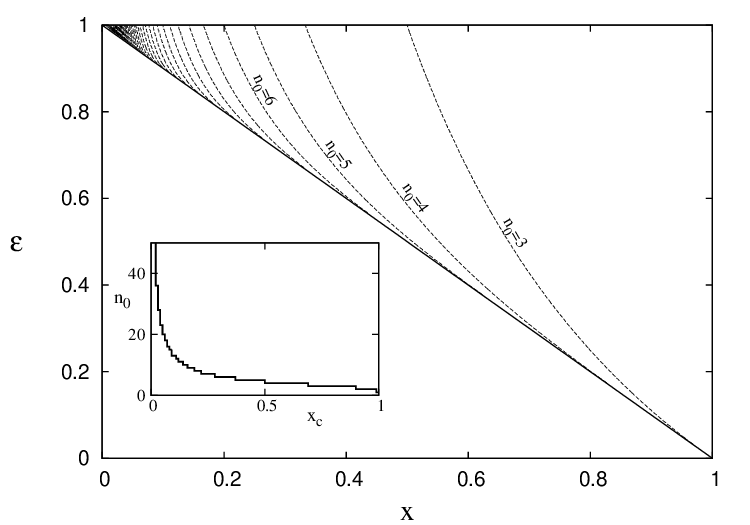} 
\par\end{centering}

\caption{\label{fig:nmap}The larger plot represents $n_{0}$ isolines as a
function of $x$ and $\epsilon$ (dashed lines) above $x_{c}(\epsilon)$
(solid line); below this line, killing all tissue cells is impossible.
The small one represents critical values $n_{0}$ in terms of $x_{c}$.}
\end{figure}

Assuming the same physical dose per fraction, $x_i=x$, as is the case
in many radiotherapy protocols, expression \eqref{eq:recursive} becomes
that of a recursive map, describing the evolution of the effective dose in
a treatment. The analysis of this map shows that, for every $\epsilon$ 
there is a critical value of $x$, 
\begin{equation}
x_{c}=1-\epsilon,\label{eq:dcrit}
\end{equation}
 dividing the plane $(\epsilon,x)$ in two different regions (see
figure \ref{fig:nmap}). For a treatment with $x<x_{c}$, there will
always be a surviving portion of the tissue since always $X_{n}<1$, for every $n$.
However, if $x>x_{c}$, after enough fractions $X_{n}>1$, meaning
that effective dose has reached the critical value and every single
cell of tissue has been removed by the treatment. Then it is possible
to find $n_{0}$, the threshold value of $n$, that kills every cell,
for a given therapy protocol. This is shown in the inset of Figure
\ref{fig:nmap}.

If the desired result is the elimination of the radiated tissue cells,
\emph{i.e.} surrounding tissue is not a concern for treatment planning,
$n_{0}$ represents the minimum number of sessions needed to achieve
this goal; any session after that will be unnecessary. On the contrary,
if the therapy goal requires the conservation of tissue cells (for
instance in order to preserve an organ), then the number of sessions
must be lower than $n_{0}$.

The session-coupling parameter $\epsilon$ is a cornerstone on isoeffect relationships.
A fractionated therapy of fully independent fractions requires a greater
radiation dose per fraction, or more fractions, in order to reach
the same isoeffect as a treatment with more correlated fractions.
The session-coupling coefficient acts here as a relaxation term. Immediately
after radiation damage occurs ($\epsilon=1$) tissue begins to recover,
as $\epsilon$ decreases, until the tissue eventually reaches its
initial radiation response capacity ($\epsilon=0$). In other words,
the formerly applied radiation results in a decrease of the annihilation
dose (initially equal to $D_{0}$) describing the effect of the next
fraction. The more coupled a session is to the previous one, the
larger the value of $\epsilon$ and, thus, the larger the effect on
the critical dose will be. Notice that unlike $\gamma$, that characterizes
the tissue primary response to radiation, $\epsilon$ characterizes
the tissue trend to recover its previous radioresistance.

Correlation between fractions can be translated in terms of the late
and acute tissue effects of radiobiology. Indeed, damaged tissue  
recovering capabilities should determine the value of $\epsilon$.
Given a dosage protocol, to an early responding tissue would correspond
$\epsilon$ close to $0$, whereas for a late responding tissue, would
be $\epsilon$ closer to $1$. Notice that in current working models
for hyperfractionated therapies this repair and recovery effects are
introduced as empirical correction factors \cite{Steel_ch8}, as
will be required for the session-coupling coefficient.

%DRP: no sé si entiendo lo que quiere decir "On the contrary, a lower
% dose per fraction brought out nonextensive properties for fractionated
% therapies.": una dosis menor por fracción hará inapreciables las
% propiedades no extensivas, al menos en esa fracción.
%OSG: La cosa es que si se quiere repartir la misma dosis total en fracciones mas
% pequeñas la no extensividad aumenta, al haber necesariamente mas fracciones
% (y mas sumas entre ellas). O lo que es lo mismo, al fraccionar mas un 
% tratamiento lo haces mas no extensivo
As it was shown in \cite{Sotolongo_prl2010}, nonextensivity properties
of tissue response to radiation for single doses are more noticeable
for higher doses than predicted by current models. On the contrary, 
for the same total dose,
a lower dose per fraction will enhance nonextensive properties in
fractionated therapies. Indeed, for high dosage a few fractions are
applied in a treatment and a change in $n$ is not required for different
$\epsilon$ values. However, in the lower dosage case, more radiation
fractions need to be applied and the $\epsilon$ parameter may become
crucial. In this case $n$ values move away from each other for isoeffect
treatments with different $\epsilon$. So, in order to achieve the
desired therapy effects, fractionated radiotherapy must be planned
for a tissue described by $\gamma$, varying $x$ according to $\epsilon$.
The session-coupling coefficient
should be experimentally studied as its
value tunes the annihilation dose along a radiotherapy protocol.

\section{Continuous formulation}

\subsection{Continuous limit}

\begin{figure}
\begin{centering}
\includegraphics[width=1\textwidth]{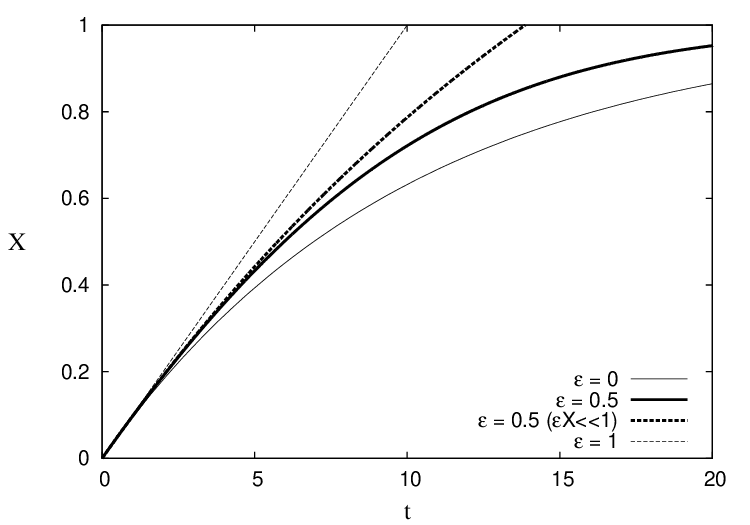} 
\par\end{centering}

\caption{\label{fig:climit}Continuous limit approximation behavior for $r=0.1$. 
As expected the solution of \eqref{eq:dx1} (thick continuous line) goes between a linear effect for 
$\epsilon = 1$ (thin continuous line) and the exponential approach to cutoff dose corresponding to
$\epsilon = 0$ (thin dashed line). Solution of \eqref{eq:dotxmix} is represented by the thick dashed line.}
\end{figure}

For some radiation treatments as brachytherapy the irradiation is
applied in a single session but for a prolonged period of time. If
the discrete irradiation sessions were close enough \eqref{eq:recursive}
could be written as,
\begin{equation}
\dot{X}=r\frac{1-X}{1-\epsilon X}\label{eq:dx1}
\end{equation}
where $r$ stands for the average absorbed radiation per unit time. 
%DRP: Una cosa que antes tampoco había visto es que, en este límite de
% fracciones casi yuxtapuestas, epsilon será prácticamente cero.
% Respecto a su relación posterior con theta, me pregunto si el límite
%  inferior de epsilon (cuando el tiempo entre fracciones tienda a cero),
% en la realidad, podría no ser nulo.
%OSG: Es al contrario, cuando el tiempo entre fracciones es cero, 
% epsilon es practicamente 1. Fijate en (14) que cuando epsilon=1 
% la dosis es aditiva, que es lo que debe ocurrir cuando son muy cercanas
At the early stages of continuous irradiation the effective dose is in general small,
and is possible to assume $\epsilon X\ll1$ and $\frac{1}{1-\epsilon X}\simeq1+\epsilon X$.
Then, 
\begin{equation}
\dot{X}\simeq r\left[1-\left(1-\epsilon\right)X\right],\label{eq:dotxmix}
\end{equation}
where the terms of second order in $\epsilon X$ and above have been neglected. 
However, as can be seen in figure \ref{fig:climit} this approximation moves 
away from \eqref{eq:dx1} as time increases.

\subsection{Continuous irradiation}

It is obvious from dose additivity properties that in the continuous
irradiation case and for two time instants $t_{0}$ and $t_{1}$ close
enough, 
\begin{equation}
X=\int_{t_{0}}^{t_{1}}rdt,\label{eq:xintr}
\end{equation}
where $r$ is the dose rate per unit time. However if both instants
of time are far enough to make relevant the tissue recovering capabilities
this expression becomes invalid. So, whereas a usual integration process
could become valid in a short time period this is not true for longer
intervals. So, in a similar way as was already done for the sum operation,
a new definition for integration must be introduced. 

This can be done following \cite{borges-2004-340} and introducing
the $q$-algebraic sum and difference,
\begin{equation}
\begin{array}{c}
x\boxplus y=x+y-\theta xy\\
x\boxminus y=\frac{x-y}{1-\theta y}
\end{array}\label{eq:qalg}
\end{equation}
where $\theta\in\left[0,1\right]$. In those terms, a nonextensive derivative operation follows such that,
\begin{equation}
\frac{{\cal D}}{dt}f=\lim_{t\rightarrow t_{0}}\frac{f\left(t\right)\boxminus f\left(t_{0}\right)}{t-t_{0}}=\frac{\dot{f}}{1-\theta f}\label{eq:formal_dif}
\end{equation}
Then we can define the physical absorbed dose rate, $r$, as the nonextensive
time derivative of the equivalent dose,

\begin{equation}
r=\frac{{\cal D}}{dt}X=\frac{\dot{X}}{1-\theta X}\label{eq:Xdif}
\end{equation}
Expression \eqref{eq:Xdif} can be rewritten as a standard ODE,
\begin{equation}
\dot{X}+\theta rX=r,\label{eq:Xode}
\end{equation}
which can be solved in the usual way taking into account that $\theta$
and $r$ are in general functions of time. In the absence of recovering effects,
the applied effective dose would increase linearly, due to the applied radiation $r$.
However a resistance
force ($\theta rX$), that depends not only on tissue recovering characteristics
but also on the dose rate and the effective dose itself, will slow
down this increase.

In order to illustrate the behaviour described by \eqref{eq:Xode},
let us suppose $r$ is
constant (a common case in clinical practice) and $\theta$ slowly
varying in time, so that it can be also taken as a constant. Then
it is straightforwardly obtained,
\begin{equation}
X=\frac{1}{\theta}\left\{ 1-\exp\left(-\theta rt\right)\right\} ,\label{eq:Xcte}
\end{equation}
allowing to find the needed irradiation time to kill every cell in
the tissue ($X=1$),
\begin{equation}
t_{k}=-\frac{\ln\left(1-\theta\right)}{\theta r},\label{eq:tk}
\end{equation}
and showing that effective dose increases at a decreasing speed,
\begin{equation}
\dot{X}=r\exp\left(-\theta rt\right),\label{eq:dotx}
\end{equation}
until tissue cells get annihilated at time $t_{k}$ ($X=1$). Under
continuous irradiation, survival fraction decreases faster at the
beginning of irradiation process. However, depending on dose rate
and $\theta$ coefficient, the killing process speed slows down until
eventually every cell is killed. If the recovery capacity is very
high ($\theta=1$) the radiation effects stack slowly and there will
always be surviving tissue cells ($t_{k}=\infty$). Those radiation
damages stack faster as long as tissue cells are less capable to recover
themselves and if there is no recovery at all ($\theta=0$) the effective
radiation dose grows linearly in time and cells get killed faster
($t_{k}=1/r$). This time shortening behavior with decreasing recovering
rate is also shown by other radiobiological models \cite{Pop1996153,huang224}.

%DRP: Y aquí es donde ese epsilon, que hemos dicho que depende del tiempo,
% haciéndose cero cuando el tiempo entre dosis sucesivas es también cero,
% parece inconsistente con el theta. Quizás no deberíamos abundar tanto
% en la relación numérica entre ambos, sino en que el papel que juegan en
% las dos teorías es el mismo... Porque la teoría discreta la deberíamos
% poder recuperar a partir de la continua, y no es así.
% Es decir, yo quitaría la última frase del último párrafo.
%OSG: OK, quitamos la ultima frase porque ahora pensandolo bien no tiene mucho sentido.
% epsilon depende del tiempo entre fracciones, en particular de la ultima
% fraccion con respecto al resto. Como si fuera una cadena de markov.
% La cosa es que al integrar en un segmento de tiempo rompemos esa cadena
% y debemos considerar los efectos de todos los instantes anteriores, no
% solo de la acumulacion tras el ultimo evento. En un tratamiento continuo,
% segun decimos aqui, theta deberia valer cerca de cero al comenzar y
% tener un comportamiento monotono creciente hasta hacerse 1 en el infinito.
% Yo me lo imagino como una arctan entre 0 y 1 (pero claro que eso es 
% como mi cabeza lo ve). Creo que es por este comportamiento que no se 
% ve la relacion directa entre epsilon y theta. La forma de relacionarlos
% seria integrar el tiempo de dos sesiones considerando un valor de r 
% promedio durante todo ese tiempo (muy bajo porque es 0 casi todo el tiempo).
% Aunque ahi no se si tendria sentido hacerlo.
% En fin que lo quitamos porque no es nada evidente.
Comparing \eqref{eq:Xode} and \eqref{eq:dotxmix} we see that, in
the limit of continuous dosage, they become the same expression with
$\theta\simeq1-\epsilon$. However this relation may become invalid
at high exposures as effective dose becomes larger and $\epsilon X$
becomes of order $1$, as shown in figure \ref{fig:climit}. 
At this point, the fractionated and continuous
treatments differ.

This shows $\theta$ could be considered constant only for a limited time of the 
continuous irradiation. It must be studied, in general, as a function of time,
describing the growing resistance of tissue to be annihilated. This function should
make that \eqref{eq:Xode} mimics the behavior of \eqref{eq:dx1}, shown in figure
\ref{fig:climit}. 
%\DRP{So $\theta$ must be studied regardless of $\epsilon$
%but if a continuous alternative therapy is desired, known $\epsilon$
%values can be a good starting point to find $\theta$.}{}

\section{Conclusions}

The use of Tsallis entropy and the maximum entropy \emph{ansatz} (second law of thermodynamics) have
allowed us to write a simple nonextensive expression for the single
dose survival fraction. The mathematical constraints, required to
define the probabilities composition such that the two limiting behaviors
are described, introduce a new parameter, relating the radiation sessions.
The fits to available experimental data show that usual treatment
have non trivial values of this parameter, \emph{i.e.}, are not close
to the limiting behaviors. This makes the study of this coefficient
relevant for clinical treatments and experimental setups. 

The existence of a varying critical dosage arises from these composition rules,
providing a criterion to adjust the critical treatment that kills every tumor cell
or minimize the damage caused to healthy tissue. This could be accomplished
changing the number of sessions or the radiation dose by session,
allowing to switch between isoeffective treatments. 

Also an expression for the effective dose in continuous irradiation
treatments has been found, showing it is phenomenologically linked
to the previous one. This has the potential to provide isoeffect relationships
in continuous dose treatments such as brachytherapy. Besides, a relation
between fractionated and continuous therapies could be established
from the obtained coefficients.

\section*{Acknowledgments}

Authors acknowledge the financial support from the Spanish Ministerio
de Ciencia e Innovación under ITRENIO project (TEC2008-06715-C02-01). 
Authors want to thanks to Juan Antonio Santos, MD for his priceless 
help in the understanding of radiobiological processes and models.

%\bibliographystyle{unsrt}
%\bibliography{osotolongo}

\begin{thebibliography}{10}

\bibitem{Swinney20041}
H.~Swinney and C.~Tsallis.
\newblock Anomalous distributions, nonlinear dynamics, and nonextensivity.
\newblock {\em Physica D: Nonlinear Phenomena}, 193(1-4):1 -- 2, 2004.

\bibitem{tsallis-1999-29}
C.~Tsallis.
\newblock Nonextensive statistics: Theoretical, experimental and computational
  evidences and connections.
\newblock {\em Brazilian Journal of Physics}, 29:1--35, 1999.

\bibitem{Sotolongo_prl2010}
O.~Sotolongo-Grau, D.~Rodr\'iguez-P\'erez, J.~C. Antoranz, and
  O.~Sotolongo-Costa.
\newblock Tissue radiation response with maximum {T}sallis entropy.
\newblock {\em Physical Review Letters}, 105(15):158105, 2010.

\bibitem{hydra1}
Jean~Paul Rieu, Arpita Upadhyaya, James~A. Glazier, Noriyuki~Bob Ouchi, and
  Yasuji Sawada.
\newblock Diffusion and deformations of single {H}ydra cells in cellular
  aggregates.
\newblock {\em Biophysical Journal}, 79(4):1903 -- 1914, 2000.

\bibitem{hydra2}
Arpita Upadhyaya, Jean-Paul Rieu, James~A. Glazier, and Yasuji Sawada.
\newblock Anomalous diffusion and non-gaussian velocity distribution of {H}ydra
  cells in cellular aggregates.
\newblock {\em Physica A: Statistical Mechanics and its Applications},
  293(3–4):549 -- 558, 2001.

\bibitem{Plastino99}
A.~Plastino and A.~R. Plastino.
\newblock Tsallis entropy and {J}aynes' information theory formalism.
\newblock {\em Brazilian Journal of Physics}, 29:50--60, 1999.

\bibitem{Sotolongo10_chamonix}
O.~Sotolongo-Grau, D.~Rodriguez-Perez, J.C. Antoranz, and O.~Sotolongo-Costa.
\newblock Non-extensive radiobiology.
\newblock In A.~Mohammad-Djafari, J-F. Bercher, and P.~Bessiere, editors, {\em
  Bayesian inference and maximum entropy methods in science and engineering
  (Proceedings of the 30th International Workshop on Bayesian Inference and
  Maximum Entropy Methods in Science and Engineering, 4-9 July 2010, Chamonix,
  France)}, volume 1305 of {\em AIP Conference Proceedings}, pages 219--226.
  AIP, 2010.

\bibitem{Plastino1994140}
A.R. Plastino and A.~Plastino.
\newblock From {G}ibbs microcanonical ensemble to {T}sallis generalized
  canonical distribution.
\newblock {\em Physics Letters A}, 193(2):140 -- 143, 1994.

\bibitem{nivanen-2003-52}
L.~Nivanen, A.~Le Mehaute, and Q.~A. Wang.
\newblock Generalized algebra within a nonextensive statistics.
\newblock {\em Reports On Mathematical Physics}, 52, 2003.

\bibitem{borges-2004-340}
E.~P. Borges.
\newblock A possible deformed algebra and calculus inspired in nonextensive
  thermostatistics.
\newblock {\em Physica A: Statistical Mechanics and its Applications}, 340:95,
  2004.

\bibitem{Kalogeropoulos2005408}
N.~Kalogeropoulos.
\newblock Algebra and calculus for {T}sallis thermo-statistics.
\newblock {\em Physica A: Statistical Mechanics and its Applications},
  356(2-4):408 -- 418, 2005.

\bibitem{kaniadakis2005}
G.~Kaniadakis, M.~Lissia, and A.~M. Scarfone.
\newblock Two-parameter deformations of logarithm, exponential, and entropy: A
  consistent framework for generalized statistical mechanics.
\newblock {\em Physical Review E}, 71, 2005.

\bibitem{isodata5}
C.~S. Parkins, J.~F. Fowler, R.~L. Maughan, and M.~J. Roper.
\newblock Repair in mouse lung for up to 20 fractions of {X} rays or neutrons.
\newblock {\em British Journal of Radiology}, 58(687):225--41, 1985.

\bibitem{isodata3}
B.~G. Douglas and J.~F. Fowler.
\newblock The effect of multiple small doses of {X} rays on skin reactions in
  the mouse and a basic interpretation.
\newblock {\em Radiation Research}, 66(2):401--426, 1976.

\bibitem{isodata1}
H.~D. Thames, R.~Withers, K.~A. Mason, and B.~O. Reid.
\newblock Dose-survival characteristics of mouse jejunal crypt cells.
\newblock {\em International Journal of Radiation Oncology*Biology*Physics},
  7(11):1591 -- 1597, 1981.

\bibitem{Steel_ch10}
A.J. van~der Kogel and C.C.R. Arnout.
\newblock Calculation of isoeffect relationships.
\newblock In G.G. Steel, editor, {\em Basic Clinical Radiobiology for Radiation
  Oncologists}, pages 72--80. Edward Arnold Publishers, London, 1993.

\bibitem{Steel_ch8}
M.~C. Joiner.
\newblock The linear-quadratic approach to fractionation.
\newblock In G.G. Steel, editor, {\em Basic Clinical Radiobiology for Radiation
  Oncologists}, pages 55--64. Edward Arnold Publishers, London, 1993.

\bibitem{Pop1996153}
L.~A.~M. Pop, J.~F. C.~M. van~den Broek, A.~G. Visser, and A.~J. van~der Kogel.
\newblock Constraints in the use of repair half times and mathematical
  modelling for the clinical application of {HDR} and {PDR} treatment schedules
  as an alternative for {LDR} brachytherapy.
\newblock {\em Radiotherapy and Oncology}, 38(2):153 -- 162, 1996.

\bibitem{huang224}
Zhibin Huang, Nina~A. Mayr, Simon~S. Lo, Jian~Z. Wang, Guang Jia, William T.~C.
  Yuh, and Roberta Johnke.
\newblock A generalized linear-quadratic model incorporating reciprocal time
  pattern of radiation damage repair.
\newblock {\em Medical Physics}, 39(1):224--230, 2012.

\end{thebibliography}

\providecommand{\noopsort}[1]{}\providecommand{\singleletter}[1]{#1}%

\end{document}